\documentclass[pre,aps,showpacs,twocolumn]{revtex4}
\usepackage{graphicx,epsfig,subfigure,amsbsy,amsmath}
\usepackage[latin1]{inputenc}

\begin{document}
\title{Quantum Mechanical Reflection Resonances}

\author{Erica Caden and  Robert Gilmore}

\affiliation{Physics Department, Drexel University,
Philadelphia,  Pennsylvania 19104, USA}

\date{\today, {\it Physical Review A}: To be submitted.}

\begin{abstract}
Resonances in the reflection probability amplitude $r(E)$
can occur in energy ranges in which the reflection
probability $R(E)=|r(E)|^2$ is 1.  They occur as the
phase $\phi(E)$ defined by
$r(E) = t^{*}(E)/t(E) = 1e^{i 2\phi(E)}$
undergoes a rapid change of $\pi$ radians.
During this transition the phase angle
exhibits a Lorentzian profile in that
$d\phi(E)/dE \simeq 1/[(E-E_0)^2+(\hbar \gamma/2)^2]$.
The energy $E_0$ identifies the location of a quasi-bound
state, $\gamma$ measures the lifetime of this state,
and $t(E)$ is a matrix element of the transfer operator.
Methods for computing and measuring these resonances are
proposed.
\end{abstract}


\maketitle

\section{Introduction}

Quantum mechanical transmission resonances, as observed for example
in the Ramsauer effect, have been known for a long time
\cite{Bohm51}. At a
transmission resonance the reflection probability is small, and
may even be zero if the potential satisfies specific properties.
Conversely, if the transmission probability is small the
reflection probability will be large.

There is a large class of potentials for which the transmission
probability is zero and therefore the reflection probability
is one in certain energy ranges.  In one dimension these potentials
have unequal asymptotic values on the left and right, with
$V_L < E < V_R$.  Here $V_L$ ($V_R$) is the asymptotic potential
on the left (right) and $E$ is the energy of a particle, incident
from the left.  One such potential is shown in Fig. 1.
This is a washboard potential,
commonly encountered with biased Josephson junctions \cite{Roberto}.
Particles incident from the left with energy $-10 < E < 19.1 eV$
will be reflected with 100\% probability, so that
$R(E) = |r(E)|^2 = 1.0$.  It is the purpose of this work to
show that the reflection probability amplitude $r(E)$
undergoes resonances in the sense
that the reflection phase shift $\phi(E)$,
defined by $r(E) = 1 e^{i 2\phi(E)}$, undergoes rapid
phase shifts when the incident energy sweeps through a
narrow energy range.  These resonances correspond
to energies at which quasistable states occur in the
potential.  The sharp resonances can be described
by a Lorentzian line shape whose peak locates the
energy of the metastable state and whose width is determined
by the lifetime of the state.

\begin{figure}[htb]   
\begin{center}
\includegraphics[angle=0,width=7.0cm]{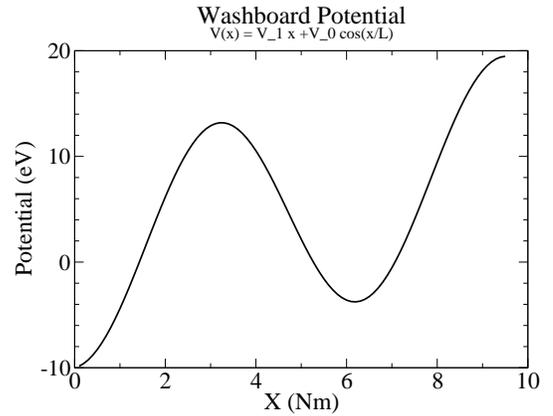}  
\end{center}
\caption{Washboard potential:
$V(x) = V_0 \cos(x/L)+V_1 x$, with $V_0=-10$ eV, $V_1 = 1$ eV, $L=1$ Nm,
$0 < x < 9.7$ Nm, and $V_L=-10$ eV, $V_R=19.9$ eV.
${\cal H} = P^2/2M+V(x)$, $M=100m_e$.  This potential
has two metastable states, at $E=0.32$ eV and $E=7.87$ eV.}
\label{washboard_potl}
\end{figure}

\section{Hamiltonian}

We choose as hamiltonian

\begin{equation}
{\cal H} = \frac{P^2}{2m}+V(\gamma)=\frac{P^2}{2M}+V_0 \cos(\gamma)+V_1
\gamma
\label{hamiltonian}
\end{equation}
Washboard potentials of this type are typically found in biased Josephson
junctions \cite{Roberto}.  In such cases the coordinate $\gamma$ is the phase
difference across the junction, $P$ is the canonically conjugate
momentum, $M=C_J \hbar^2/4e^2$ is expressed in terms of the
junction capacitance $C_J$ and fundamental constants $e$ and $\hbar$,
$V_0$ is the Josephson energy, and $V_1/V_0=I/I_c$,
where $I$ is the bias current and $I_c$ is the critical current.
For simplicity, we will interpret this hamiltonian as describing
a particle of mass $M = 100 m_e$ with coordinate $x=\gamma L$
in the potential whose parameters are as described in the
caption of Fig. \ref{washboard_potl}.

\section{Reflection Phase}

The phase shift can be computed by constructing the
transfer matrix for the potential.
This relates the input and output amplitudes on the left
with those on the right (the notation is as used in \cite{Gil04}):

\begin{equation}
\left[  \begin{array}{c}
A_L \\ B_L \end{array} \right] =
\left[  \begin{array}{cc}
t_{11}(E) & t_{12}(E) \\ t_{21}(E) & t_{22}(E) \end{array} \right]
\left[  \begin{array}{c}
A_R \\ B_R \end{array} \right]
\end{equation}
The reflection amplitude is $r(E) = t_{21}(E)/t_{11}(E)$.
For reflecting boundary conditions these two matrix elements
are complex conjugates of each other:

\begin{equation}
t_{21}(E) = t_{11}(E)^{*}~~~~~~~~~    t_{11}(E) = a(E) +ib(E)
\end{equation}
In the neighborhood of a metastable state the
the matrix element $t_{11}(E)$ has the form
$e^{-i \theta}[(E-E_0)-i(\hbar \gamma/2)]$,
where the rapidly varying phase
$\tan^{-1}\left( \hbar \gamma/2/(E-E_0)\right)$
rides on top of a slowly varying $\theta(E)$.
The total phase  is
$\phi(E) = \theta(E) +
\tan^{-1}\left( (\hbar \gamma/2)/(E-E_0)\right)$.
From this we immediately derive a Lorentzian
line shape:

\begin{equation}
\frac{d\phi(E)}{dE}-\frac{d\theta(E)}{dE} =
\frac{(\hbar \gamma/2)}{(E-E_0)^2 + (\hbar \gamma/2)^2}
\end{equation}
When $\theta(E)$ is slowly varying over the narrow
range of a resonance, the term $d\theta/dE$ can be neglected
and $d\phi(E)/dE$ shows a Lorentzian lineshape.

\begin{figure}[htb]   
\begin{center}
\includegraphics[angle=0,width=7.0cm]{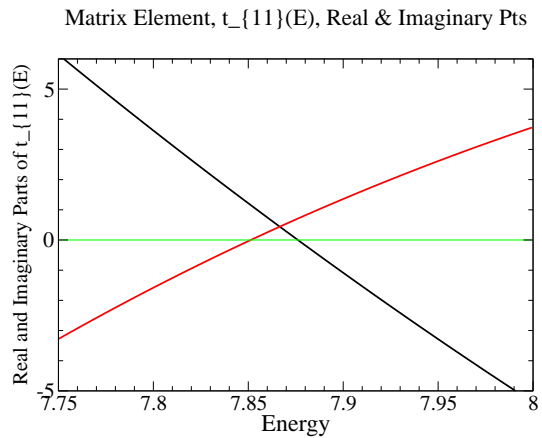}  
\end{center}
\caption{Real (negative slope) and imaginary
(positive slope) parts $a(E)$ and $b(E)$
of the matrix element $t_{11}(E)=a(E)+ib(E)$.}
  \label{real_imag}
\end{figure}

\begin{figure}[htb]   
\begin{center}
\includegraphics[angle=0,width=7.0cm]{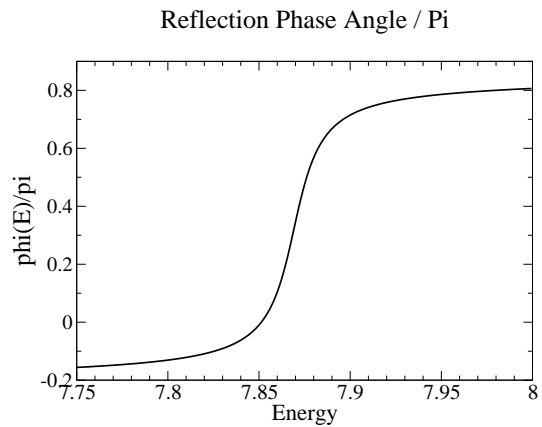}  
\end{center}
\caption{Phase shift $\phi(E)/\pi$ computed from the
real and imaginary parts of the matrix element $t_{11}(E)$.}
\label{phase_angle}
\end{figure}

The real and imaginary parts of the matrix element
$t_{11}(E)$ have been computed for the potential
shown in Fig. \ref{washboard_potl}.
They are plotted in the neighborhood of the first excited resonance
at $E = +7.87$ eV in Fig. \ref{real_imag}.  The phase shift $\phi(E)$
computed from the real and imaginary parts of $t_{11}(E)$
are shown in Fig. \ref{phase_angle}.

\begin{figure}[htb]   
\begin{center}
\includegraphics[angle=0,width=7.0cm]{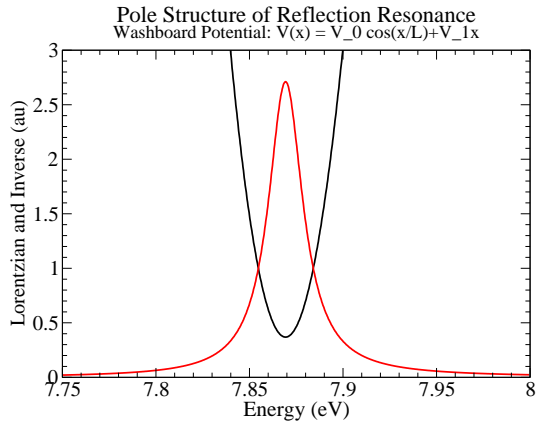}  
\end{center}
\caption{Reflection resonance for the
washboard potential shown in Fig. 1.
Also shown is its inverse, which should be a simple
parabola if the resonance is a Lorentzian.}
\label{lorentzian}
\end{figure}

The real and imaginary parts of these matrix elements
can be written in the form

\begin{equation}  \begin{array}{rcl}
a(E) &=& \alpha(E) (E-E_r)\\
b(E) &=& \beta(E) (E-E_i)  \end{array}
\end{equation}
They have zero crossings at $E_r$ and $E_i$:
the sharper the resonance, the closer the crossings.
The slopes $\alpha(E)$ and $\beta(E)$ are related to
$\cos(\theta)$ and $\sin(\theta)$.  Generally the slopes
are only  weakly dependent on $E$.  When this is the case

\begin{equation}\begin{array}{c}
|t_{11}(E)|^2 = |t_{21}(E)|^2 = a(E)^2+b(E)^2 =\\
   \\
(\alpha^2 + \beta^2) [(E-E_0)^2 + (\hbar \gamma/2)^2]\end{array}
\end{equation}
Its inverse has a Lorentzian shape
\begin{equation}
\frac{1}{|t_{11}(E)|^2} = \frac{\rm cst}{ (E-E_0)^2 +(\hbar \gamma/2)^2}
\end{equation}
The center $E_0$ and halfwidth $\hbar \gamma/2$ of the peak are weighted
averages of the zero crossings of the real and imaginary parts
of $t_{11}(E)$:

\begin{equation}
E_0 = \frac{\alpha^2 E_r + \beta^2 E_i}{\alpha^2  + \beta^2 }~~~~~~~~
\hbar \gamma/2 = \frac{|\alpha \beta (E_r-E_i) |}{\alpha^2  + \beta^2 }
\end{equation}
Figure \ref{lorentzian} plots the Lorentzian $1/|t_{11}(E)|^2$ in the
neighborhood of a sharp resonance for the washboard potential.
Also plotted is the inverse of this function, which should be
a simple parabola if the peak does in fact have a Lorentzian
profile.  When the slowly varying part of $\phi(E)$
is removed, so that $\phi(E)$ varies rapidly
through $\pi$ radians as $E$ passes the resonance, a plot of
$d\phi(E)/dE$ vs. $E$ produces a Lorentzian line shape
centered at the resonance.  The plots of $1/|t_{11}(E)|^2$ and
$d\phi(E)/dE$ are indistinguishable.

\section{Physical Interpretation}

Rapid phase changes in transmission and/or reflection
amplitudes are associated with the existence of
metastable states.  They have also received a nice
interpretation by Wigner as time delays caused by
these resonances \cite{wigner}.  That is, as a particle enters
a region where a metastable state can exist, it undergoes
multiple internal reflections before exiting this
region.  These internal reflections are responsible for
the prolonged time delay before transmission or reflection.
The Wigner time delay is measured quantitatively by
$\delta t = \hbar d\phi/dE$.  This delay time has
been studied under scattering conditions \cite{96},
reflecting boundary conditions \cite{97}, and
metastable escape conditions \cite{98,99}.

The phase shift calculated in Sect. III shows Lorentzian
peaks near resonances riding on top of a slowly varying
background.  The slowly varying background is due to the
slight time delay for the reflected particle due to
the longer distance it must travel before reaching the
classically forbidden region.  The fast variation is the
excess time delay spent in the well near the energy
of the metastable (resonant) state.

\section{Differential Spectroscopy}

\begin{figure}[htb]   
\begin{center}
\includegraphics[angle=0,width=7.0cm]{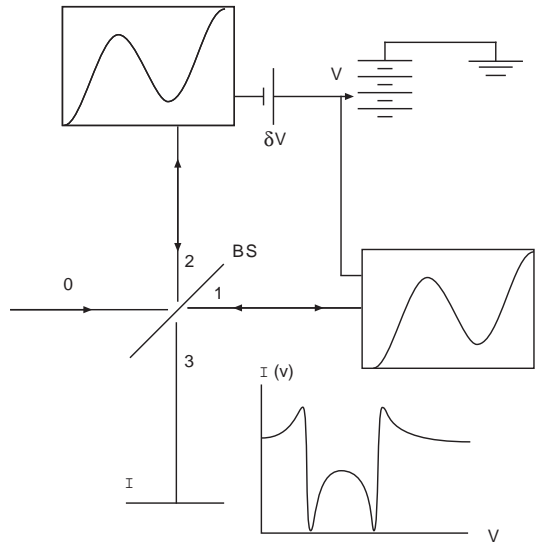}
\end{center}
\caption{Reflection resonance interferometer.
A beam of thermal electrons $E \simeq 0$ is incident
on a beamsplitter.  The two resulting beams are reflected
off identical washboard potentials that are biased at
$-V$ and $-V-\delta V$, with $\delta V$ of the order of 10
halfwidths of the resonant peak.  The bias potential
$V$ is scanned through the resonance and the intensity
of the recombined beam in the output arm is monitored.}
\label{interferometer}
\end{figure}

It may not always be easy to extract phase information
from reflection amplitudes.  An alternative method for locating
reflection resonances would therefore be useful.  One
such method is illustrated in Fig. \ref{interferometer}.
This is a schematic for a Differential Reflection Resonance
Spectrometer.  A beam of thermal electrons ($E \simeq 0$ eV)
is incident on a beam splitter from arm 0 of the interferometer.
The beam is split into arms 1 and 2.  Each of these beams is
reflected off identical washboard potentials.  The two
reflecting potentials are biased at voltages $-V$ and
$-V-\delta V$ \cite{Gil04}.  The separation
$\delta V$ between the two biasing
potentials is chosen small, of the order of 10 profile halfwidths.
The potential $V$ is scanned through the resonance, and the
intensity profile of the interfering beams in arm 3 is
recorded.  The split beams undergo rapid $2\pi$ phase
shifts approximately $\delta V$ apart.  As these two phase
shifts occur the output intensity undergoes a rapid
symmetric variation as shown in Fig. \ref{int_pattern}.
The symmetry of this pattern and the separation of the
rapidly fluctuating features by $\delta V$ is a clear
indication that a reflection resonance exists approximately
where the first feature undergoes its most rapid variation.
This result is insensitive to amplitude and phase inequalities
in the two returning beams due to different arm lengths and
unequal splitting in the beam splitter.

\section{Lorentz Profiles from Intensities}

\begin{figure}[htb]   
\begin{center}
\includegraphics[angle=0,width=7.0cm]{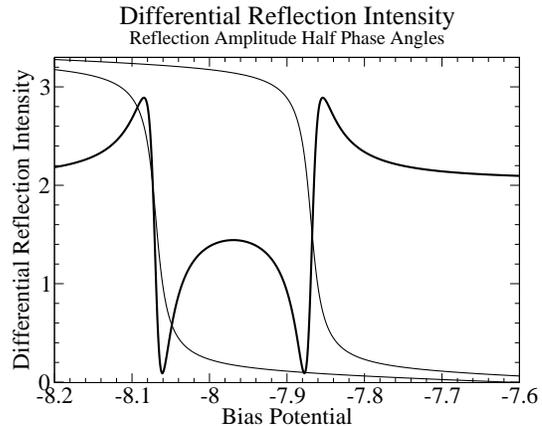}
\end{center}
\caption{Intensity in the output arm as a function of
biasing voltage $V$ in
the Differential Reflection Resonance Spectroscopy
setup shown in Fig. 5 (darker curve).  For this simulation we used
$\delta V=0.2$ eV,
$r_1=e^{i 2\phi(V)}$ and $r_2 = a e^{i \theta} e^{i 2\phi(V+\delta V)}$
with $a = 0.7$ and $\theta = 0.2 \times 2\pi$.
The plots of $\phi(V)$ and $\phi(V+\delta V)$ are shown
in lighter curves after being shifted down by an equal phase.}
\label{int_pattern}
\end{figure}

Line shape information can be extracted from intensity
information.  The total amplitude of the signal in
the measurement arm of the interferometer is

\begin{equation}
A(V) = a_1e^{i \alpha_1} r_1(V)+a_2e^{i \alpha_2} r_2(V+\delta V)
\end{equation}
Here $r_1(V)=e^{i 2\phi(V)}$ is the reflection amplitude for
particles reflected from the potential biased
at $V$ in arm 1, and similarly for
$r_2(V+\delta V)=e^{i 2\phi(V+\delta V)}$
in arm 2.  The returning beams are weighted with
complex amplitudes $a_ie^{i \alpha_i}$, with $a_i > 0$.
The amplitudes $a_i$ are typically unequal due to
nonperfect beam splitting, and the difference
in the angles $\alpha_i$ is related to differences in
arm lengths.  The observed intensity is

\begin{equation}\begin{array}{c}
I(V)=|A(V)|^2 = a_1^2+a_2^2+\\
  \\
2a_1a_2 \cos \left[ (\alpha_1-\alpha_2)
+2(\phi(V)-\phi(V+\delta V)) \right]\end{array}
\end{equation}
A typical intensity profile is shown in Fig. \ref{int_pattern}.
For this pattern $\delta V = 0.2eV$,
$a_1=1, \alpha_1=0$ and $a_2=0.7, \alpha = 0.2
\times 2\pi$.
The intensity varies between $I_{\rm Max}$ and $I_{\rm min}$,
where $\cos \left[* \right]$ assumes values $+1$ and $-1$.
Several features of this profile are worth mentioning.
(1)  The pattern is symmetric about the midpoint $V+\frac{1}{2}\delta V$.
This comes about because the intensity is a function of the
difference $2(\phi(V)-\phi(V+\delta V))$, and these
two functions are identical except for the energy shift.  Outside
the range of rapid variation this difference is zero, while
on transiting the resonance region the difference changes
from 0 to $2 \pi$ radians, then back down to 0.
(2)  The pattern has five critical points.  The two critical points
near the resonance at $-V=-7.9$ eV occur as $2\phi(V)$ changes
through $2\pi$ radians and the two critial points near $-V-\delta V$
occur as $2\phi(V+\delta V)$ changes through $2\pi$ radians.
The fifth critical point, at the symmetry point
$V+\frac{1}{2} \delta V$, is due to the overlap of the
shoulders of the two Lorentzians centered near $V=-7.9$ eV and
$V=-8.1$ eV, as will be shown shortly.

\begin{figure}[htb]   
\begin{center}
\includegraphics[angle=0,width=7.0cm]{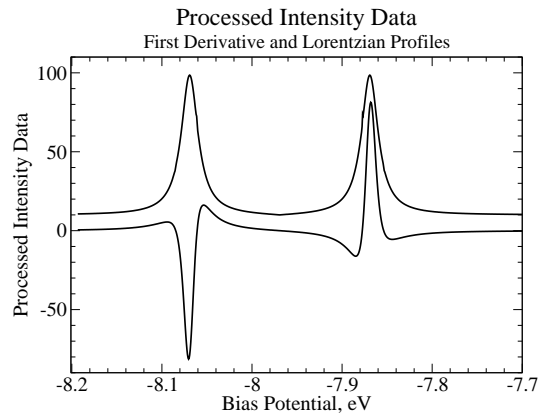}
\end{center}
\caption{Plot of $dI/dV$ for the intensity shown in Fig.
\ref{int_pattern}.  Vertically offset: plot of the
function $dI/dV/\sqrt{(I_{\rm Max}-I)(I-I_{\rm min})}$,
showing recovery of two Lorentzian line profiles
at the appropriate biasing potentials.}
\label{processed}
\end{figure}

The derivative of the intensity is

\begin{equation}\begin{array}{c}
\displaystyle  \frac{dI}{dV} = -2a_1a_2 \sin\left[ (\alpha_1-\alpha_2)
+2(\phi(V)-\phi(V+\delta V)) \right]\\
   \\
\displaystyle
\times 2
\left(  \frac{d\phi(V)}{dV} -\frac{d\phi(V+\delta V)}{dV}
\right)\end{array}
\end{equation}
At $I_{\rm Max}$ and $I_{\rm min}$ where $\cos \left[* \right]=\pm1$,
$\sin \left[ * \right]$ passes through zero.  As a result
$2a_1a_2\sin \left[ *\right] = \pm \sqrt{(I_{\rm Max}-I)(I-I_{\rm min})}$.
Therefore, the intensity data can be processed according to

\begin{equation}
\pm \frac{dI(V)/dV}{\sqrt{(I_{\rm Max}-I)(I-I_{\rm min})}}=
2\left(  \frac{d\phi(V)}{dV} -\frac{d\phi(V+\delta V)}{dV}
\right)
\end{equation}
The right hand side is the difference of two Lorentzians.
This difference vanishes at the midpoint $V+\frac{1}{2} \delta V$,
where the falling shoulder of the Lorentian around the
resonance near $-7.9$ eV is equal to the negative of the rising shoulder of
the displaced resonance near $-8.1$ eV
(fifth critical point of $dI(V)/dV$).
The expression on the left of Equ.(11) is difficult to compute
near the critical points where $I=I_{\rm Max}$ or $I_{\rm min}$.
In these regions the intensity should be approximated
in second order, and the singular expression
simplifies to
$\sqrt{|d^2I(V)/dV^2|}{\sqrt{(I_{\rm Max}-I_{\rm min})/2}}$.
The absolute value of the right hand side of Equ.(11)
is plotted in Fig. \ref{processed} for the intensity data
shown in Fig. \ref{int_pattern}.  The absolute value consists
of two Lorentzian profiles, one surrounding the resonance
near $-7.9$ eV, the other around the displaced resonance
near $-8.1$ eV.  These two peaks are displaced vertically
from the curve $dI(V)/dV$, shown below it.

\section{Conclusions}

For scattering potentials possessing bound states,
the phase of the matrix element $t_{11}(E)$ increases
by $\pi$ radians each time $E$ increases through a bound
state value (real pole of the $S$-matrix) and increases
rapidly through $\pi$ radians each time $E$ passes through
a transmission resonance.  For such resonances the
transmission line shape $1/|t_{11}(E)|^2$ is well
approximated by a Lorentzian.  For reflecting potentials
of the type discussed here (c.f., Fig. \ref{washboard_potl}),
the reflection amplitude
$r(E)= t_{11}(E)^{*}/t_{11}(E)$ increases rapidly through
$2\pi$ radians each time $E$ passes through
a reflection resonance.  For such resonances the
line shape $1/|t_{11}(E)|^2$ is well approximated by a Lorentzian.
The Lorentzian can be computed from the real and imaginary
parts of the appropriate transfer or scattering matrix
element (c.f., Equ.(5) and Fig. \ref{lorentzian}).
It can also be computed as $d\phi(E)/dE$
(c.f., Fig. \ref{phase_angle}).  We
have described how this information can be extracted
from unique intensity signatures obtained from a
Differential Reflection Resonance Interferometer
(c.f., Equ. (11) and Figs. \ref{int_pattern} and \ref{processed}).
Quasibound states in washboard potentials are currently
located experimentally by spectroscopic methods \cite{Roberto}.
This requires at least one of the states to be occupied.
The method proposed here provides a completely different
approach for locating such states.  Further,
the states need not be occupied.

\end{document}